# ASSESSING THE EFFECTIVENESS OF CURRENT CYBERSECURITY REGULATIONS AND POLICIES IN THE US

Ejiofor Oluomachi, Akinsola Ahmed, Wahab Ahmed, Edozie Samson,

*Abstract-* This article assesses the effectiveness of current cybersecurity regulations and policies in the United States amidst the escalating frequency and sophistication of cyber threats. The focus is on the comprehensive framework established by the U.S. government, with a spotlight on the National Institute of Standards and Technology (NIST) Cybersecurity Framework and key regulations such as HIPAA, GLBA, FISMA, CISA, CCPA, and the DOD Cybersecurity Maturity Model Certification. The study evaluates the impact of these regulations on different sectors and analyzes trends in cybercrime data from 2000 to 2022. The findings highlight the challenges, successes, and the need for continuous adaptation in the face of evolving cyber threats.

*Index Terms*- Cybersecurity, Regulations, Policies, NIST Cybersecurity Framework, HIPAA, GLBA, FISMA, CISA, CCPA, DOD CMMC, Cyber Threats, Data Compromises, Fraud, Trend Analysis, Public-Private Partnerships, Cyber Resilience.

## I. INTRODUCTION

Cybersecurity in the United States has become a critical and dynamic field, given the increasing frequency and sophistication of cyber threats. The nation's reliance on digital infrastructure for communication, commerce, and critical services has made it a prime target for cyberattacks. Government agencies, private enterprises, and individuals are all vulnerable to a wide range of cyber threats, including data breaches, ransomware attacks, and nation-state-sponsored cyber espionage. As a response to these challenges, the United States has developed a comprehensive cybersecurity framework that involves collaboration between government, industry, and academia to enhance the resilience of the nation's cyberspace [1].

One of the key pillars of the U.S. cybersecurity strategy is the National Institute of Standards and Technology (NIST) Cybersecurity Framework. This framework provides a set of best practices and guidelines for organizations to manage and improve their cybersecurity posture. It emphasizes risk management, continuous monitoring, and incident response to enhance overall cybersecurity resilience. Additionally, various federal agencies, such as the Department of Homeland Security (DHS) and the Federal Bureau of Investigation (FBI), play crucial roles in protecting critical infrastructure and investigating cybercrimes.

The U.S. Cyber Command, established in 2009, is responsible for defending the nation against significant cyber threats and ensuring the military's cyber capabilities [2].

Despite these efforts, challenges persist, and the cybersecurity landscape is constantly evolving. The private sector also plays a vital role in cybersecurity, with industries investing heavily in technologies and personnel to secure their networks and data. Ongoing collaboration between the public and private sectors, international cooperation, and a commitment to staying ahead of emerging threats are essential components of the United States' approach to cybersecurity. As threats continue to evolve, policymakers and cybersecurity professionals must remain vigilant and adaptive to safeguard the nation's digital infrastructure effectively [3].

The United States has witnessed a concerning rise in cyber threats and attacks, reflecting an alarming trend of increasing cybercriminal activities. As technology advances, so does the sophistication of cyber threats, posing significant challenges to the country's cybersecurity landscape. A study published in the "Association of Computing Machinery" highlights the surge in ransomware attacks, where malicious actors encrypt sensitive data and demand payment for its release. These attacks have targeted various sectors, including healthcare, finance, and critical infrastructure, causing disruptions and financial losses. The study emphasizes the need for continuous improvement in cybersecurity measures to mitigate the evolving tactics employed by cybercriminals [3].

Furthermore, nation-state-sponsored cyber espionage has become a growing concern for the United States. A comprehensive report by the "Cybersecurity and Infrastructure Security Agency (CISA)" highlights the persistent and sophisticated cyber threats posed by nation-states seeking to compromise U.S. government networks and critical infrastructure. The report underscores the importance of enhancing cyber defenses and information-sharing mechanisms to safeguard national security. The increasing frequency and severity of cyber threats underscore the urgency for a holistic and collaborative approach involving government agencies, private industries, and academia to fortify the nation's cybersecurity posture [4].

In the United States, cybersecurity laws and regulations play a crucial role in safeguarding sensitive information and critical infrastructure from cyber threats like hacking, malware, and data breaches. They establish a framework for organizations to secure their networks and systems effectively. These legal measures also serve to hold companies and individuals accountable for any cyber incidents, ensuring that victims of cybercrime have legal recourse.

Overall, these laws create a foundational standard for protecting sensitive information and critical infrastructure from cyber threats. It is important to recognize that certain laws and regulations may be industry-specific, and compliance can vary depending on the circumstances.

## II. LITERATURE REVIEW

### 2.1 Key Cybersecurity Regulations and Policies in the U.S.

**NIST Cybersecurity Framework:**

The National Institute of Standards and Technology (NIST) developed the Cybersecurity Framework to provide organizations with guidelines and best practices to manage and improve their cybersecurity risk management. The framework includes standards, guidelines, and practices for enhancing cybersecurity. The latest version is NIST Cybersecurity Framework 1.1, released in April 2018 [5]

The National Institute of Standards and Technology (NIST) Cybersecurity Framework in the United States is a set of guidelines and best practices designed to help organizations manage and improve their cybersecurity risk management processes. Developed by NIST in response to Executive Order 13636, the framework provides a flexible and voluntary framework that organizations across various sectors can adopt to enhance their cybersecurity resilience. The framework is based on five core functions: Identify, Protect, Detect, Respond, and Recover. Each function represents key aspects of a comprehensive cybersecurity program, and organizations can use the framework to assess and strengthen their cybersecurity posture.

The first function, "Identify," focuses on understanding and managing cybersecurity risks by establishing an organizational context, understanding asset management, and identifying risk management strategies. The "Protect" function involves implementing safeguards to ensure the delivery of critical infrastructure services, including access controls, training, and data protection measures. The "Detect" function emphasizes continuous monitoring and timely identification of cybersecurity events, while the "Respond" function outlines actions to take in the event of a cybersecurity incident. The final function, "Recover," involves developing and implementing strategies for the timely restoration of services and systems following a cybersecurity incident.

Organizations in the United States are encouraged to use the NIST Cybersecurity Framework to assess and improve their cybersecurity posture, aligning with their risk management needs and business objectives. The framework's adaptability allows it to be applied across diverse sectors, contributing to a more resilient and secure cyberspace.

The NIST Cybersecurity Framework (CSF) is a comprehensive and adaptive set of guidelines and best practices designed to assist organizations in managing and enhancing their cybersecurity risk management processes. Developed by the National Institute of Standards and Technology (NIST), the framework provides a flexible approach that organizations across various sectors can use to strengthen their cybersecurity posture. The framework is structured around three primary components: the Core, the Implementation Tiers, and the Framework Profile.

**Core Functions:**

The Core of the NIST CSF consists of five functions, each representing a key aspect of an effective cybersecurity program:

**Identify:** This function involves understanding and managing cybersecurity risks by establishing an organizational context, conducting asset management, and developing risk management strategies.

**Protect:** The Protect function focuses on implementing safeguards to ensure the reliable delivery of critical infrastructure services. This includes measures such as access controls, training and awareness programs, and data protection measures.

**Detect:** Detect emphasizes continuous monitoring and timely identification of cybersecurity events. This involves implementing processes to identify anomalies and events that could indicate a cybersecurity incident.

**Respond:** In the event of a cybersecurity incident, organizations must have a robust response plan. The Respond function outlines actions to take promptly to mitigate the impact of incidents.

**Recover:** The Recover function involves developing and implementing strategies to restore services and systems affected by a cybersecurity incident. This includes prioritizing recovery efforts and incorporating lessons learned for future improvements.

**Implementation Tiers:**

The Implementation Tiers provide a way for organizations to characterize their approach to managing cybersecurity risk. There are four tiers, ranging from Partial (Tier 1) to Adaptive (Tier 4), reflecting the organization's maturity and sophistication in implementing the framework's functions.

**Framework Profile:**

The Framework Profile is a customizable set of cybersecurity outcomes aligned with the organization's business needs and risk management priorities. It enables organizations to tailor the use of the framework to their specific circumstances, ensuring that cybersecurity efforts align with overall business objectives.

**HIPAA (Health Insurance Portability and Accountability Act):**

HIPAA establishes standards for the protection of sensitive patient health information. Covered entities, including healthcare providers and health insurance companies, must comply with HIPAA regulations to ensure the confidentiality, integrity, and availability of health information [6].

The Health Insurance Portability and Accountability Act (HIPAA) in the United States is a critical piece of legislation enacted in 1996 to safeguard the privacy and security of individuals' health information. HIPAA addresses the challenges associated with the electronic transmission of healthcare data and aims to ensure the confidentiality and integrity of patients' sensitive information. The act consists of multiple rules, with the Privacy Rule and the Security Rule being the two primary components governing the protection of health information. The Privacy Rule establishes standards for the use and disclosure of protected health information (PHI), while the Security Rule focuses on the security safeguards necessary to protect electronic PHI.

The HIPAA Security Rule outlines a comprehensive framework for securing electronic PHI (ePHI). It mandates that covered entities, such as healthcare providers, health plans, and healthcare clearinghouses, implement administrative, physical, and technical safeguards to ensure the confidentiality, integrity, and availability of ePHI. The framework includes measures such as access controls, encryption, audit controls, and risk assessments. Covered entities are required to conduct regular risk assessments to identify and mitigate potential vulnerabilities in their information systems and adopt policies and procedures to safeguard ePHI against unauthorized access or disclosure.

Non-compliance with HIPAA regulations can result in severe penalties, including fines and legal actions. The framework provided by HIPAA is essential for promoting trust in the healthcare system and protecting individuals' sensitive health information in an increasingly digital and interconnected healthcare environment [6].

**Gramm-Leach-Bliley Act (GLBA):**

GLBA requires financial institutions to safeguard customers' private financial information. It includes provisions for the security and confidentiality of nonpublic personal information and mandates the development and implementation of information security programs [7].

The Gramm-Leach-Bliley Act (GLBA), also known as the Financial Modernization Act of 1999, is a landmark piece of legislation in the United States aimed at enhancing consumer privacy and information security in the financial sector. Enacted to repeal certain provisions of the Glass-Steagall Act, GLBA introduced new requirements for financial institutions, such as banks, credit unions, and securities firms, regarding the protection of consumers' nonpublic personal information (NPI). The primary focus of GLBA is to ensure that financial institutions establish appropriate safeguards to protect the confidentiality and security of customer information (U.S. Congress, 1999) [8].

Under GLBA, financial institutions are required to develop, implement, and maintain comprehensive information security programs. These programs must include administrative, technical, and physical safeguards to protect the security, confidentiality, and integrity of customer information. The Act also mandates that financial institutions provide consumers with clear and concise privacy notices, detailing the institution's information-sharing practices and giving customers the option to opt-out of having their information shared with non-affiliated third parties.

Non-compliance with GLBA can result in significant penalties and regulatory actions. The Act underscores the importance of transparency in information-sharing practices and aims to empower consumers with control over how their personal financial information is used and shared by financial institutions [9].

**Federal Information Security Modernization Act (FISMA)**

The Federal Information Security Modernization Act (FISMA), enacted in 2002, is a key cybersecurity legislation in the United States. It mandates federal agencies to implement security controls to safeguard their information systems and data. The primary goal is to ensure the confidentiality, integrity, and availability of the information collected, stored, and utilized by federal agencies [10].

FISMA requires agencies to establish comprehensive information security programs, encompassing regular risk assessments, security testing, incident response planning, and continuous monitoring of security controls. Additionally, agencies are required to report their compliance with the law to both the Office of Management and Budget (OMB) and the Department of Homeland Security (DHS).

FISMA designates the National Institute of Standards and Technology (NIST) as the principal entity tasked with creating security standards and guidelines for federal agencies. NIST's "NIST Special Publication 800-53" provides a comprehensive framework detailing the security controls that federal agencies must adopt to meet FISMA requirements.

**Cybersecurity Information Sharing Act (CISA)**

The Cybersecurity Information Sharing Act (CISA), enacted in 2015 by the U.S. Congress, encourages private companies to share information on cyber threats with the government and grants liability protections for such sharing. Its primary goal is to enhance the exchange of cyber threat information between the government and the private sector, aiming to safeguard critical infrastructure and national security from cyber-attacks. CISA permits private companies to share cyber threat data with federal agencies, including the Department of Homeland Security (DHS), and allows reciprocal sharing of information from the government to private entities.

Additionally, CISA includes provisions for establishing voluntary Information Sharing and Analysis Organizations (ISAOs). These organizations facilitate the sharing of cyber threat information among their members. The law offers liability protections to companies sharing information in good faith to combat cyber threats. However, critics, including privacy and civil liberties advocates, argue that CISA may lack sufficient safeguards for personal information and could potentially be exploited for government surveillance (Enterprise Engineering Solutions, 2023) [11].

**California Consumer Privacy Act (CCPA):**

Although primarily focused on privacy, the CCPA includes requirements for businesses to implement reasonable security measures to protect consumers' personal information. The California Consumer Privacy Act (CCPA) is a landmark privacy law in the United States that grants California residents' greater control over their personal information held by businesses. Enacted in 2018 and effective from January 1, 2020, the CCPA empowers consumers by providing them with the right to know what personal information is collected, sold, or disclosed by businesses. Covered businesses, which include those with annual

                                                                                     

gross revenues over $25 million or those that handle personal information of at least 50,000 California consumers, are required to disclose their data practices and honor consumer requests for access, deletion, or opt-out of the sale of their personal information.

The CCPA introduces a comprehensive framework for privacy protection, promoting transparency and accountability in the collection and use of personal data. It requires businesses to update their privacy policies, inform consumers about their rights, and establish mechanisms for consumers to exercise those rights. Additionally, the law imposes strict obligations on businesses to implement reasonable security practices to safeguard consumer information. The CCPA has significant implications not only for businesses operating in California but also for the broader privacy landscape in the United States, influencing discussions around potential federal privacy legislation [12].

**DOD Cybersecurity Maturity Model Certification (CMMC):**

The Department of Defense (DOD) introduced the CMMC to enhance the protection of Controlled Unclassified Information (CUI) within the defense industrial base. Contractors and suppliers must meet specific cybersecurity standards to participate in DOD contracts. The Department of Defense (DOD) Cybersecurity Maturity Model Certification (CMMC) in the United States is a comprehensive framework designed to enhance cybersecurity practices within the defense industrial base (DIB). Introduced to address concerns surrounding the protection of sensitive information in the supply chain, CMMC establishes a five-level maturity model, ranging from basic cyber hygiene to advanced capabilities. Each level builds on the previous one, encompassing specific domains and capabilities critical to cybersecurity. Notably, CMMC represents a departure from self-assessment to third-party certification, ensuring a more rigorous and standardized evaluation process. The DOD plans to gradually implement CMMC requirements into new contracts by the mid-2020s, emphasizing collaboration with industry experts and stakeholders to adapt to evolving cyber threats. Challenges include potential implementation costs for organizations within the DIB, especially smaller companies, and the ongoing need for updates to address the dynamic nature of the cybersecurity threat landscape [13].

**EFFECTIVENESS OF CURRENT CYBERSECURITY REGULATIONS AND POLICIES IN THE US**

Cybersecurity regulations and policies play a crucial role in safeguarding the digital landscape of the United States. In recent years, the increasing frequency and sophistication of cyber threats have prompted the government to establish a comprehensive framework to protect critical infrastructure, sensitive data, and national security. The effectiveness of these regulations and policies is a subject of ongoing scrutiny and evaluation.

One significant initiative in the realm of cybersecurity is the Cybersecurity Enhancement Act of 2014, which reinforced research and development efforts, established best practices and enhanced the coordination between the government and private sector. The act emphasized the importance of information sharing and collaboration to mitigate cyber threats effectively. However, the effectiveness of these measures relies on the active participation and cooperation of various stakeholders, including government agencies, private enterprises, and individuals.

The establishment of the Cybersecurity and Infrastructure Security Agency (CISA) in 2018 marked a milestone in consolidating cybersecurity efforts within the U.S. government. CISA plays a central role in coordinating cybersecurity initiatives, providing support to critical infrastructure entities, and disseminating threat intelligence. The agency's efforts are aimed at fortifying the nation's resilience against cyber threats, but challenges such as resource constraints and evolving threat landscapes persist.

Moreover, sector-specific regulations, such as the Health Insurance Portability and Accountability Act (HIPAA) for the healthcare sector and the Payment Card Industry Data Security Standard (PCI DSS) for the financial industry, contribute to a more tailored approach in addressing sector-specific vulnerabilities. These regulations mandate specific cybersecurity measures and practices, fostering a proactive stance against cyber threats within industries that handle sensitive information.

However, the effectiveness of these regulations is contingent on their enforcement and adaptability to rapidly evolving cyber threats. The lack of a one-size-fits-all solution poses challenges, as industries have unique cybersecurity requirements and face distinct threats. Continuous updates and amendments to existing regulations are necessary to address emerging risks and technological advancements.

Public-private partnerships are integral to the success of cybersecurity regulations in the U.S. The government and private sector collaboration enhances information sharing, incident response capabilities, and overall cyber resilience. Initiatives such as the National Institute of Standards and Technology (NIST) Cybersecurity Framework provide a flexible and risk-based approach that organizations can adopt to bolster their cybersecurity posture. However, voluntary adoption of these frameworks may result in uneven implementation across different sectors.

The 2015 cyberattack on the Office of Personnel Management (OPM), which exposed sensitive personal information of millions of federal employees. This incident underscored the vulnerabilities within government agencies and led to a heightened focus on improving cybersecurity practices. In response, the federal government took steps to strengthen its cybersecurity policies, emphasizing the importance of robust defense mechanisms and proactive risk management to safeguard sensitive data (U.S. Office of Personnel Management, 2015).

The financial sector has also faced significant cyber threats, with breaches targeting credit card information and customer data. The 2014 data breach at JPMorgan Chase is a notable case. The incident prompted increased scrutiny on the financial industry's cybersecurity practices and spurred discussions on the need for more stringent regulations. Subsequently, financial institutions have had to comply with regulatory frameworks such as the PCI DSS to enhance the security of payment card transactions [14].

In the healthcare sector, the WannaCry ransomware attack in 2017 impacted organizations globally, including the National Health Service (NHS) in the UK. While not directly targeting the U.S., the incident raised concerns about the potential vulnerabilities in healthcare systems. In response, the U.S. government and healthcare organizations reinforced efforts to comply with and strengthen regulations like HIPAA to protect patient information and enhance overall cybersecurity resilience [15].

The SolarWinds supply chain attack, discovered in late 2020, exemplifies the challenges in defending against sophisticated cyber threats. The incident compromised multiple U.S. government agencies and private organizations by exploiting a trusted software vendor. This case underscores the need for comprehensive cybersecurity policies that address supply chain risks and highlights the importance of continuous monitoring and threat intelligence sharing [16]

Despite these challenges, success stories also exist. The implementation of the NIST Cybersecurity Framework has positively impacted organizations across various sectors. The framework's adoption by companies like Microsoft and its integration into industry standards demonstrates its effectiveness in providing a flexible and risk-based approach to cybersecurity. Cybersecurity regulations and policies in the U.S. have made significant strides in enhancing the nation's cyber resilience. The establishment of key agencies, sector-specific regulations, and collaborative frameworks reflects a proactive approach to addressing cyber threats. However, the dynamic nature of cybersecurity challenges necessitates ongoing evaluation and adaptation of these measures to ensure their continued effectiveness. A comprehensive, collaborative, and adaptive approach is crucial for staying ahead of evolving cyber threats and protecting the nation's critical assets.

### III. METHODOLOGY

**Data Collection and Sources**

Trend analysis will be conducted using time series data collected over the years at different ranges. These secondary data were extracted from the Bureau of Economic Analysis and United States Databases. The dataset spans 2000 and 2022. The study considers the following indicators, including cost of cybercrimes (in millions) Data Compromises, number of records exposed in millions, Fraud (in million dollars). Charts were employed to show trends and patterns displayed by the indicators.

### IV. RESULTS

Following the adopted methodology, the analyses are discussed in detail:

**Cost of Cyber Crimes**

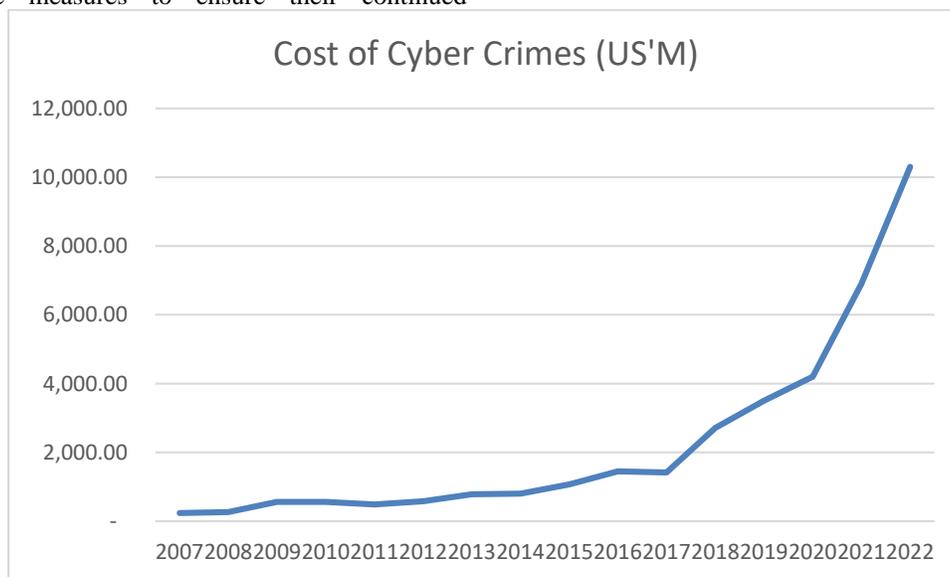

The trend analysis of the "Cost of Cyber Crimes" dataset spanning from 2007 to 2022 reveals a substantial and consistent increase in financial losses attributed to cyber threats over the examined period. The data portrays a clear escalation from $239.10 million in 2007 to a staggering $10,300.00 million in 2022. This upward trajectory indicates a significant and concerning surge in the financial impact of cybercrimes on various entities.

Noteworthy is the period from 2010 onwards, where the costs experienced a pronounced increase. This escalation suggests a critical turning point in the landscape of cyber threats, signifying a more aggressive and damaging nature of cyber-attacks. The year 2022 stands out as a peak, representing the highest financial toll recorded in the dataset.

While the year-over-year breakdown exhibits varying degrees of increase, certain years witness more substantial spikes, indicating potential shifts in the methods or intensity of cyber-attacks during those periods. This nuanced analysis allows for a more granular understanding of the trends, facilitating a targeted approach in cybersecurity efforts. The consistent surge in the cost of cybercrimes highlights the urgent need for robust cybersecurity measures.

**Data Compromises**

   

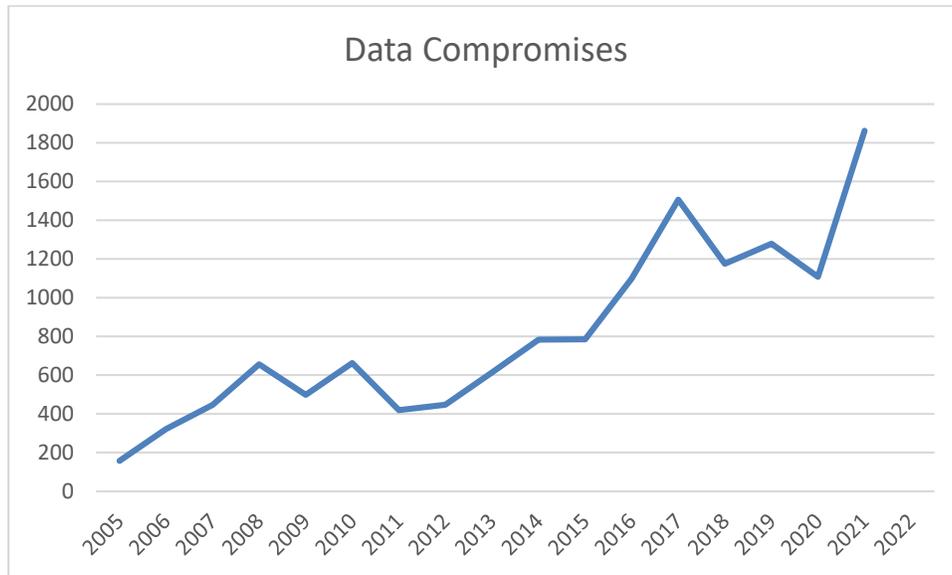

The trend analysis of the "Data Compromises" dataset from 2005 to 2022 uncovers interesting patterns in the frequency of reported occurrences. The data illustrates a progressive increase in data compromises, starting with a sudden leap from 157 instances in 2005 to 662 in 2010. In the following years, there were oscillations in the number of incidents, with a peak of 1862 occurrences in 2021. It is important to consider the lack of data for 2022 when analyzing the trend.

Starting from 2015, there has been a continuous increase in data breaches, which suggests a heightened attention and increased complexity of cyber-attacks. In the dataset, the year 2021 is notable for having the largest number of reported data intrusions. This underscores the expanding range of potential dangers and the necessity for increased cybersecurity measures to safeguard confidential data.

Examining the year-on-year fluctuations offers valuable insights into possible alterations in cyber-attack tactics. The variations in the frequency of data breaches in certain years may suggest alterations in cybercriminal strategies or preferences for targets. Organizations and cybersecurity professionals should carefully examine these tendencies to adjust their protection measures properly.

**Fraud**

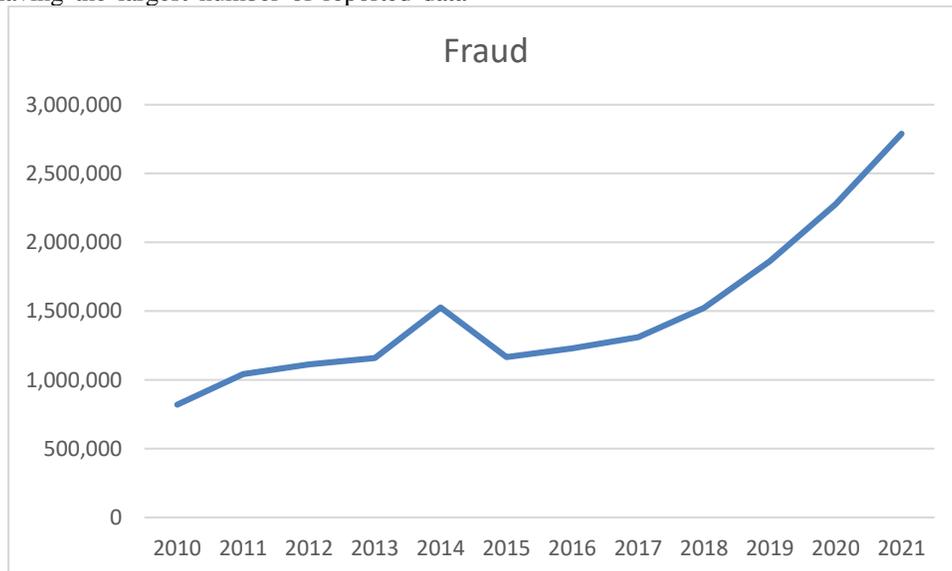

The trend analysis of the "Fraud" dataset covering the years 2010 to 2021 demonstrates fluctuations and consistent increases in reported fraud cases. In 2010, the recorded fraud instances stood at 820,072, with a subsequent rise in the following years. Notably, the dataset reveals a surge in reported fraud cases from 2014 onwards, reaching a peak of 2,789,161 cases in 2021.

The years 2014 to 2016 witnessed a significant spike in reported fraud, with the numbers escalating from 1,526,365 in 2014 to 1,228,865 in 2016. While there was a slight decrease in 2015, the overall trend points to an increasing prevalence of fraud

        

during this period. The data also highlights the noteworthy impact of the COVID-19 pandemic, as demonstrated by the substantial jump in reported fraud cases from 1,862,871 in 2019 to 2,277,130 in 2020. This surge may be attributed to the changing dynamics of the global landscape, with cybercriminals exploiting vulnerabilities arising from the pandemic.

**Summary Statistics of Variables**
**Table 1: Descriptive Analysis**

| Variables | Minimum | Maximum | Mean | Std. Deviation |
|---|---|---|---|---|
| **Cost of Cyber Crimes (US'M)** | 207.39 | 10300.00 | 2013.94 | 2720.96 |
| **Data Compromises** | 157.00 | 1862.00 | 853.39 | 478.171 |
| **Fraud (USD)** | 820072.00 | 2789161.00 | 1553312.87 | 598004.75 |

*Source: Author's Computation (2024) Using Stata 14*

Table 1 shows that the average cost of cybercrimes measured in billions of US dollars is $2,013.94 billion with a standard deviation of $2,720.96 billion indicating a notable variability in reported cost of cybercrime. The minimum cost of cybercrime is $207.39 billion, while the maximum cost of cybercrime is $10,300 billion. The average value of data compromise is 853.39 instances with a standard deviation of 478.17 instances. The minimum value of data compromise is 157 instances, while the maximum value of data compromises is 1,862 instances. The average value of fraud is $1,553,312.87 billion, with a standard deviation of $598,004.75. The deviation of $598,004.75 reflects the degree of dispersion in reported fraud amounts, suggesting variability in the financial impact of fraud incidents. The minimum value of fraud is $820,072 billion, while the maximum value of fraud is $2,789,161 billion.

## V. DISCUSSION OF FINDINGS

The wide range in the "Cost of Cyber Crimes" suggests a significant financial impact on entities operating in the United States. It was discovered that there was substantial variability in reported costs, emphasizing the diverse nature of cyber threats. it is crucial to recognize that some incidents incur much higher costs, underscoring the severity of certain cyberattacks. This finding implies that the current cybersecurity landscape demands a multifaceted approach to regulation, addressing a spectrum of threats. The varying number of data compromises highlights the frequency and scope of incidents where sensitive information is compromised. This finding emphasizes the need for regulations that not only prevent data breaches but also mitigate their impact when they occur. The effectiveness of current policies can be assessed by how well they address the root causes of data compromises and contribute to reducing their occurrence and severity. The study also found diversity in the financial impact of fraudulent activities. Effective cybersecurity policies should not only target the prevention of cybercrimes but also focus on minimizing the economic losses resulting from fraudulent activities. The analysis suggests that policies should address emerging trends in cyber fraud, adapting to evolving techniques employed by threat actors.

The results emphasize the intricate and ever-changing nature of cybersecurity difficulties encountered by organizations in the United States. The efficacy of existing cybersecurity legislation can be assessed based on their capacity to adjust to the changing threat environment, thoroughly tackle various cyber threats, and minimize the financial and operational repercussions of cyber incidents. Policymakers should consistently assess and improve current policies to ensure they stay strong and adaptable to the constantly evolving cybersecurity landscape.

## REFERENCES

[1]   [1] U.S. Department of Homeland Security (2024). (https://www.dhs.gov/)
[2]   [2] National Institute of Standards and Technology. (https://www.nist.gov/cyberframework)
[3]   [3] The Evolving Menace of Ransomware: A Comparative Analysis of Pre-pandemic and Mid-pandemic Attacks Michael Lang https://dl.acm.org/doi/full/10.1145/3558006
[4]   [4] U.S. Cyber Command (2024). (https://www.cybercom.mil/)
[5]   [4] Cybersecurity & Infrastructure Security Agency (CISA). (2020). SolarWinds Supply Chain Attack: "Active Exploitation of SolarWinds Software." Retrieved from https://www.cisa.gov/active-exploitation-SolarWinds-software
[6]   [5] National Institute of Standards and Technology. (2018). National Institute of Standards and Technology (NIST) Cybersecurity Framework: "Framework for Improving Critical Infrastructure Cybersecurity." Retrieved from https://www.nist.gov/cyberframework
[7]   [6] U.S. Department of Health & Human Services. (2003). Health Insurance Portability and Accountability Act (HIPAA) Security Rule. Retrieved from https://www.hhs.gov/sites/default/files/hipaa-simplification-201303.pdf
[8]   [7] Privacy of Consumer Financial Information Rule Under the Gramm-Leach-Bliley Act retrieved through https://www.federalregister.gov/documents/2019/04/04/2019-06039/privacy-of-consumer-financial-information-rule-under-the-gramm-leach-bliley-act
[9]   [8] The Gramm-Leach-Bliley Act and Financial Integration retrieved from https://www.federalreservehistory.org/essays/gramm-leach-bliley-act
[10]   [9] Privacy of Consumer Financial Information (Regulation S-P) retrieved from https://www.sec.gov/rules/2000/06/privacy-consumer-financial-information-regulation-s-p#:~:text=Under%20the%20Gramm%2DLeach%2DBliley,to%20the%20consumer%20and%20the
[11]   [10] Federal Information Security Modernization Act of 2014 (FISMA 2014) retrieved from https://www.cisa.gov/topics/cyber-threats-and-advisories/federal-information-security-modernization-act
[12]   [11] Enterprise Engineering Solutions. (2023). Cybersecurity Laws And Regulations In US (2023). Retrieved from https://www.eescorporation.com/cybersecurity-laws-and-regulations-in-us/
[13]   [12] California Legislative Information. (2018). California Consumer Privacy Act (CCPA). Retrieved from https://leginfo.legislature.ca.gov/faces/billTextClient.xhtml?bill_id=201720180AB375
[14]   [13] U.S. Department of Defense. (2020). CMMC Frequently Asked Questions. Retrieved from https://www.acq.osd.mil/cmmc/faq.html
[15]   [14] Department of Justice. (2015). JPMorgan Chase Data Breach: "Russian National Pleads Guilty to Largest Data Breach Conspiracy Ever Charged in the United States." Retrieved from

https://www.justice.gov/opa/pr/russian-national-pleads-guilty-largest-data-breach-conspiracy-ever-charged-united-states

[16] [15] NHS ransomware attack spreads worldwide Roger Collier 10.1503/cmaj.1095434

[17] [16] Cybersecurity & Infrastructure Security Agency (CISA). (2020). SolarWinds Supply Chain Attack: "Active Exploitation of SolarWinds Software." Retrieved from https://www.cisa.gov/active-exploitation-SolarWinds-software

AUTHORS

**First Author** – Ejiofor Oluomachi, Department of Computer Science, Austin Peay State University, Clarksville USA., Oejiofor@my.apsu.edu

**Second Author** – Akinsola Ahmed, Department of Computer Science, Austin Peay State University, Clarksville USA., aakinsola@my.apsu.edu

**Third Author** – Wahab Ahmed, Department of Computer Science, Austin Peay State University, Clarksville USA., Awahab@my.apsu.edu

**Fourth Author** – Edozie Samson, Department of Computer Science, Austin Peay State University, Clarksville USA., Sedozie@my.apsu.edu